\documentclass{spie}

\usepackage{graphicx}
\usepackage{amsmath}
\usepackage{amssymb}

\newcommand {\intset}     {\mathbb{Z}}
\newcommand {\ket}[1]     {|{#1}\rangle}

\newcommand {\setof}[1]   {\left\{{#1}\right\}}
\newcommand {\sqmatrix}[4] {\left( \begin{array}{cc}
                                    {#1} & {#2} \\ 
				    {#3} & {#4} 
				    \end{array}
		            \right)}

\newcommand {\onesqrt}   {{1 \over \sqrt 2}}
\newcommand {\onesq}[1]  {{1 \over \sqrt {#1}}}
\newcommand {\oneof}[1]  {{1 \over {#1}}}
\newcommand {\epr}       {{\onesqrt(\ket{00}+\ket{11})}}

\newcommand {\BFA}	{\mbox{BFA}}
\newcommand {\BHA}	{\mbox{BHA}}
\newcommand {\FA}	{\mbox{FA}}
\newcommand {\HA}	{\mbox{HA}}
\newcommand {\XAN}	{\mbox{XAN}}
\newcommand {\AN}	{\mbox{AN}}
\newcommand {\COPY}	{\mbox{COPY}}
\newcommand {\SWAP}	{\mbox{\em SWAP}}

\title{Distributed quantum computing: A distributed Shor algorithm}
\author{Anocha Yimsiriwattana and Samuel J. Lomonaco Jr.\supit{a}
   \skiplinehalf
   \supit{a}Dept. of Computer Sci. and Electrical Eng., University of Maryland,
      Baltimore County,\\ 1000 Hilltop Circle, Baltimore, MD, USA, 25000.
}
\authorinfo{
   Further author information: (Send correspondence to Anocha Yimsiriwattana)\\
   Anocha Yimsiriwattana: 
      E-mail: {\tt ayimsi1@umbc.edu}, 
      URL: {\tt http:/userpages.umbc.edu/\symbol{126}ayimsi1}\\
   Samuel J. Lomonaco Jr.:
      E-mail: {\tt lomonaco@umbc.edu}, 
      URL: {\tt http:/www.cs.umbc.edu/\symbol{126}lomonaco}
}

\begin{document}
\maketitle

%\baselineskip 0.7cm
%%%%%%%%%%%%%%%%%%%%%%%%%%%%%%%%%%%%%%%%%%%%%%%%%%%%%%%%%%%%%%%%%%%%%%%%%%%%%%%
\begin{abstract}

We present a distributed implementation of Shor's quantum factoring algorithm
on a distributed quantum network model. This model provides a means for small
capacity quantum computers to work together in such a way as to simulate a
large capacity quantum computer. In this paper, entanglement is used as a
resource for implementing non-local operations between two or more quantum
computers. These non-local operations are used to implement a distributed
factoring circuit with polynomially many gates. This distributed version of
Shor's algorithm requires an additional overhead of $O((\log N)^2)$
communication complexity, where $N$ denotes the integer to be factored.

\end{abstract}

\keywords{Shor's algorithm, factoring algorithm, distributed quantum algorithms,
quantum circuit.}
%%%%%%%%%%%%%%%%%%%%%%%%%%%%%%%%%%%%%%%%%%%%%%%%%%%%%%%%%%%%%%%%%%%%%%%%%%%%%%%

%%%%%%%%%%%%%%%%%%%%%%%%%%%%%%%%%%%%%%%%%%%%%%%%%%%%%%%%%%%%%%%%%%%%%%%%%%%%%
\section{Introduction}
%%%%%%%%%%%%%%%%%%%%%%%%%%%%%%%%%%%%%%%%%%%%%%%%%%%%%%%%%%%%%%%%%%%%%%%%%%%%%
To utilize the full power of quantum computation, one needs a scalable quantum
computer with a sufficient number of qubits. Unfortunately, the first practical
quantum computers are likely to have only small qubit capacity. One way to
overcome this difficulty is by using the distributed computing paradigm.  By a
distributed quantum computer, we mean a network of limited capacity quantum
computers connected via classical and quantum channels. Quantum entangled
states, in particular generalized GHZ states, provide an effective way of
implementing non-local operations, such as, non-local CNOTs and
teleportation~\cite{eisert:non-local,anocha:distqc}.

We use distributed quantum computing techniques to construct a distributed
quantum circuit for the Shor factoring algorithm. Let $n=\log N$, where $N$ is
the number to be factored. The gate complexity of this particular distributed
implementation of Shor's algorithm is $O(n^3)$ with $O(n^2)$ communication
overhead.\footnote{Shor's factoring algorithm is of gate complexity $O(n^2 \log
n \log\log n)$ and space complexity $O(n\log n\log\log n)$.}

In section \ref{sec:distqc}, the general principles of distributed quantum
computing are outlined, and two primitive distributed computing operators, {\bf
cat-entangler} and {\bf cat-disentangler}, are introduced. We use these two
primitive operators to implement non-local operations, such as non-local CNOTs
and teleportation. Then we discuss how to share the cost of implementing a
non-local controlled $U$, where $U$ can be decomposed into a number of gates.
The section ends with an distributed implementation of the Fourier transform.

In section \ref{sec:factoring}, we give a detailed description of an
implementation of Shor's non-distributed factoring algorithm. This
implementation is based on the phase estimation and order finding algorithms.
We discuss in detail how to implement ``modular exponentiation,'' which
implementation will be used later in this paper as a blueprint for creating a
distributed quantum algorithm.

In section \ref{sec:distshor}, we implement a distributed factoring algorithm
by partitioning the qubits into groups in such a way that each group fits on
one of the computers making up the network. We then proceed to replace
controlled gates with non-local controlled gates whenever necessary.

%%%%%%%%%%%%%%%%%%%%%%%%%%%%%%%%%%%%%%%%%%%%%%%%%%%%%%%%%%%%%%%%%%%%%%%%%%%%%
\section{Distributed quantum computing}
\label{sec:distqc}
%%%%%%%%%%%%%%%%%%%%%%%%%%%%%%%%%%%%%%%%%%%%%%%%%%%%%%%%%%%%%%%%%%%%%%%%%%%%%
By a distributed quantum computer (DQC), we mean a network of limited capacity
quantum computers connected via classical and quantum channels. Each computer
(or node) possesses a quantum register that can hold only a fixed limited
number of qubits. Each node also possesses a small fixed number of channel
qubits which can be sent back and forth over the network. Each register qubit
can freely interact with any other qubit within the same register. Each such
qubit can also freely interact with channel qubits that are in the same
computer. In particular, each such qubit can interact with other qubits on a
remote computer by two methods: 1) The qubit can interact via non-local
operations, or 2) The qubit can be teleported or physically transported to a
remote computer in order to locally interact with a qubit on that remote
computer.

Indeed, distributed quantum computing can be implemented by only teleporting or
physically transporting qubits back and forth. However, a more efficient
implementation of DQC has been proposed by Eisert et al~\cite{eisert:non-local}
using non-local CNOT gates. Since the controlled-NOT gate together with all
one-qubit gates is universal set of gates~\cite{shor:factoring}, a distributed
implementation of any unitary transformation reduces to the implementation of
non-local CNOT gates.  Eisert et al also prove that one shared entangled pair
and two classical bits are necessary and sufficient to implement a non-local
CNOT gate.

\begin{figure}[h]
\begin{center}
   \includegraphics{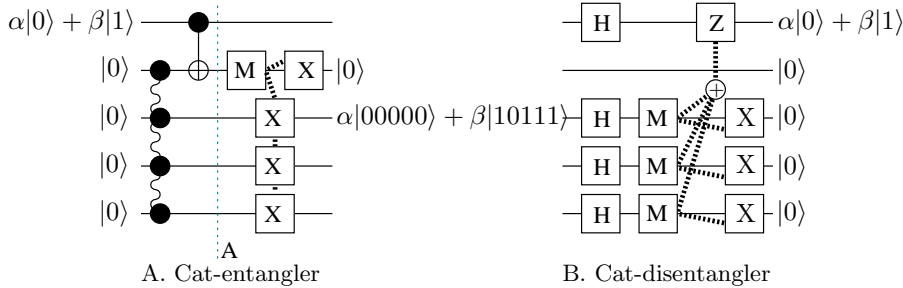}
   \caption{\label{fig:entdis} 
      The cat-entangler and cat-disentangler operations for a 5-qubit system
      are shown in figure \ref{fig:entdis}-A and figure \ref{fig:entdis}-B,
      respectively. A dotted-line represents a measurement result, which is
      classical and is used to control $X$ gates. The $Z$ gate in circuit
      \ref{fig:entdis}-B is controlled by the exclusive-or $(\oplus)$ of the
      three classical bits resulting from the measurement of qubits three to
      five. A qubit is reset to $\ket{0}$ by a control-$X$ gate. This
      control-$X$ gate is controlled by a classical bit arising from the
      measurement on the qubit.
   }
\end{center}
\end{figure}

Yimsiriwattana and Lomonaco \cite{anocha:distqc} have identified two primitive
operations, {\bf cat-entangler} and {\bf cat-disentangler}, which can be used
to implement non-local operations, e.g., non-local CNOTs, non-local controlled
gates, and teleportation.  Figure~\ref{fig:entdis} illustrates cat-entangler
and cat-disentangler operations. 

For the implementation of a non-local CNOT gate, an entangled pair must first
be established between two computers. Then, the cat-entangler is used to
transform a control qubit $\alpha \ket{0} + \beta \ket{1}$ and an entangled
pair $\epr$ into the state $\alpha \ket{00} + \beta \ket{11}$, called a
``{\bf cat-like}'' state. This state permits two computers to share the
control qubit.  As a result, each computer now can use a qubit shared within
the cat-like state as a local control qubit. 

\begin{figure}[h]
\begin{center}
   \includegraphics{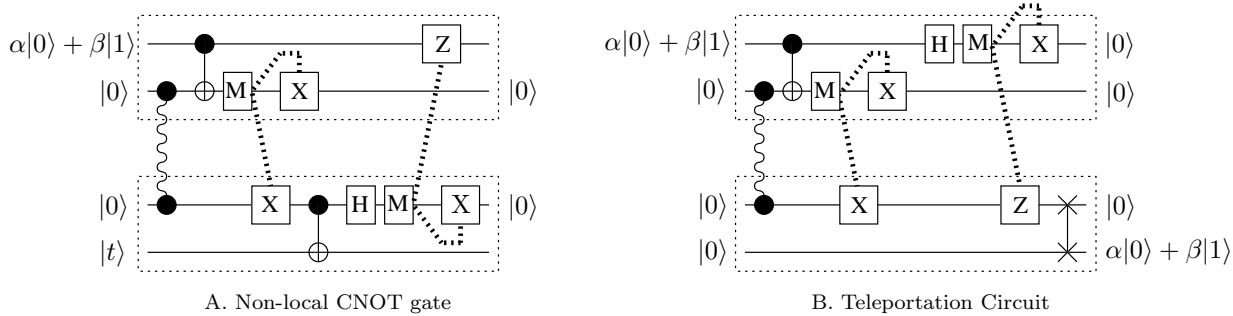}
   \caption{\label{fig:cnotteleport} 
      This figure shows both the non-local CNOT (A) and the teleportation
      circuits (B). In both circuits, the cat-entangler creates a cat-like
      state, which is shared between the first and the third qubits.  In the
      non-local CNOT circuit (A), the third line shares with the first line the
      same control qubit via the cat-like state. It is used as a local control
      qubit to control the target qubit. Finally the cat-disentangler is
      applied to disentangle the control qubit from the cat-like state and
      return the control qubit back to the first line. In the teleportation
      circuit, the cat-disentangler disentangles the unknown qubit from the
      cat-like state, and transfers the unknown state to the third qubit.
   }
\end{center}
\end{figure}

After completion of the control operation, the cat-disentangler is then applied
to disentangle and restore the control qubit from the cat-like state.  Finally,
channel qubits are reset by using the classical information resulting from
measurement to control $X$ gates. In this way, channel qubits can be reused
and entangled pairs can be re-established. A non-local CNOT circuit is illustrated in figure~\ref{fig:cnotteleport}-A.

To teleport an unknown qubit from computer A to B, we begin by establishing an
entangled pair between two computers. Then, we apply the cat-entangler
operation to create a cat-like state from an unknown qubit and the entangled
pair. After that, we apply a cat-disentangler operation to disentangle and
restore the unknown qubit from the cat-like state into the computer B.
Finally, we reset the channel qubits by swapping the unknown qubit with
$\ket{0}$. The teleportation circuit is shown in figure
\ref{fig:cnotteleport}-B.  

Because a cat-like state permits two computers to share a control qubit, the
cost of implementing a non-local controlled $U$, where $U$ is a unitary
transformation composed of a number of basic gates, can be shared among these
basic gates.  

\begin{figure}[h]
\begin{center}
   \includegraphics{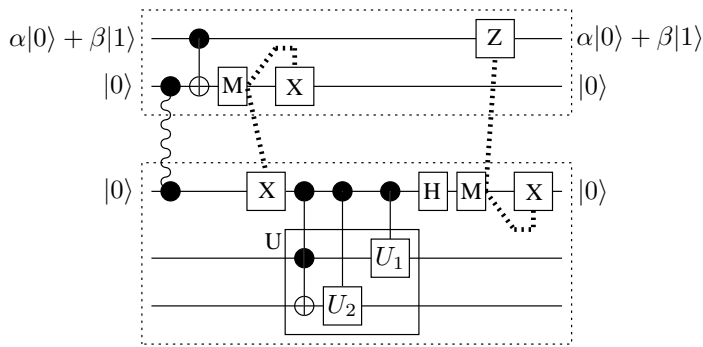}
   \caption{\label{fig:nonlocal-cu} 
      Assume $U = U_1\cdot U_2\cdot CNOT$.  Then a controlled $U$ can be
      distributed as shown.  The control line needs to be distributed only
      once, because it can be reused. This implementation allows the cost of
      distributing the control qubit to be shared among the elementary gates.
   }
\end{center}
\end{figure}

For example, let us assume that a unitary transformation has the form $U =
U_1\cdot U_2\cdot U_3$, where $U_3=CNOT$. Since the control qubit is reused,
each non-locally controlled $U_i$ gate can be implemented using asymptotically
only $\oneof{3}$ entangled pair and $2 \over 3$ classical bit, as demonstrated
in figure~\ref{fig:nonlocal-cu}.

Before the execution of a non-local operation, an entangled pair must first be
established between channel computers. If each machine possesses two channel
qubits, then two entangled pairs can be established by sending two qubits.  To
do so, each computer begins by entangling its own channel qubits, then
exchanging one qubit of the pair with the other computer.  As a result, one
entangled pair is established at the asymptotically cost of sending one
qubit. To refresh the entanglement, the procedure is simply repeated after the
channel qubits are reset to the state $\ket{0}$.

\subsection{Distributed quantum Fourier transform}
%%%%%%%%%%%%%%%%%%%%%%%%%%%%%%%%%%%%%%%%%%%%%%%%%%%%%%%%%%%%%%%%%%%%%%%%%%%%%
%
\begin{figure*}
   \begin{center}
   \includegraphics{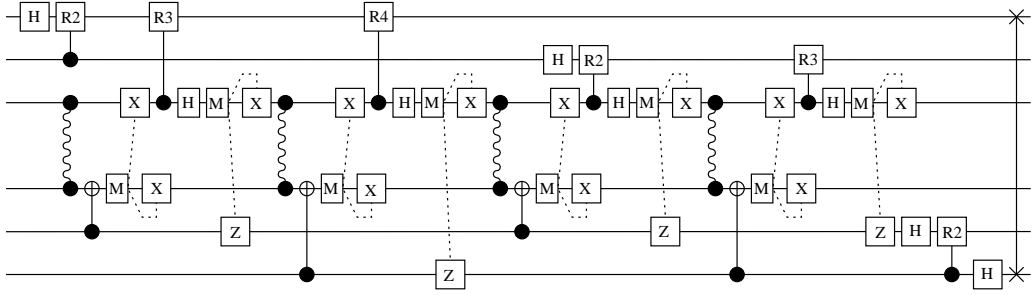}
   \caption{\label{fig:dist-fourier} 
      This figure shows an implementation of distributed quantum Fourier
      transform for 4 qubits, implemented on two machines, using non-local
      $R_k$ gates. The swap gate can be implemented by teleporting qubits back
      and forth between two computers.
   }
   \end{center}
\end{figure*}
The quantum Fourier transform is a unitary transformation defined on standard
basis states as follows,
\begin{equation}
   \label{eqn:fourier}
   \ket{j} \rightarrow \oneof{\sqrt{2^n}}
                       \sum_{k=0}^{2^n-1} e^{2\pi \imath jk/2^n}\ket{k},
\end{equation}
where $n$ is the number of qubits. 

An efficient circuit for the quantum Fourier transformation can be found in
Nielsen and Chuang's book~\cite{neilsen:qbook} and also in Cleve et al. paper
\cite{cleve:qal-revisited}. We implement a distributed version of the Fourier
transformation by replacing a controlled $R_k$ with non-local controlled $R_k$,
when necessary.  The distributed swap gate can be implemented by teleporting
qubits back and forth between two computers.  An implementation of the
distributed Fourier transformation of $4$ qubits is shown in
figure~\ref{fig:dist-fourier}, where the gate $R_k$ is defined as:
\begin{eqnarray} 
   R_k = \sqmatrix{1}{0}{0}{e^{2\pi \imath/2^k}}, 
\end{eqnarray}
for $k \in \setof{2,3,\ldots}$.  

For a more detailed discussion on distributed quantum computing, please consult
Yimsiriwattana and Lomonaco \cite{anocha:distqc}.

\section{The quantum factoring algorithm}
\label{sec:factoring}
%%%%%%%%%%%%%%%%%%%%%%%%%%%%%%%%%%%%%%%%%%%%%%%%%%%%%%%%%%%%%%%%%%%%%%%%%%%%%%
{\bf The prime factorization problem} is defined as follows: {\em Given a
composite odd positive number $N$, find its prime factors}
\cite{lomonaco:factoring}.

It is well known that factoring a composite number $N$ reduces to the task of
choosing a random integer $a$ relatively prime to $N$, and then determining its
multiplicative order $r$ modulo $N$, i.e., to find the smallest positive
integer $r$ such that $a^r = 1\ (mod\ N)$. This problem is known as the ``order
finding problem.''

Cleve et al~\cite{cleve:qal-revisited} have shown that the order finding
problem reduces to the phase estimation problem, a problem which can be solved
efficiently by a quantum computer. We briefly review these problems in this
section.

\subsection{Phase Estimation Algorithm}
%%%%%%%%%%%%%%%%%%%%%%%%%%%%%%%%%%%%%%%%%%%%%%%%%%%%%%%%%%%%%%%%%%%%%%%%%%%%%
{\bf The phase estimation problem} is defined as follows: {\em Let $U$ be an
$n$-qubit unitary transformation having eigenvalues
\[ 
   \lambda_0 = e^{2\pi \imath \theta_0}, \ldots, \lambda_{2^n-1} = e^{2\pi
   \imath \theta_{2^n-1}} 
\]
with corresponding eigenkets
\[
   \ket{\psi_0},\ldots, \ket{\psi_{2^n-1}}
\]
where $0 \le \theta_k < 1$.  Given one of the eigenket $\ket{\psi_t}$,
estimates the value of $\theta_t$.}

Cleve et al solve this problem as follows: Construct two quantum registers, the
first an $m$-qubit register, and the second an $n$-qubit register. Then
construct a unitary transformation $c_m(U)$ which acts on both registers as
follows:
\begin{equation}
   c_m(U):\ket{k}\ket{\psi} \rightarrow \ket{k}U^k\ket{\psi}
\end{equation}
where $\ket{k}$ and $\ket{\psi}$ denotes respectively the state of the first and
second register. The phase estimation algorithm can be described as follows: 

{\tt
\noindent {\bf Phase Estimation Algorithm:} \\
{\bf Input}: $U$ and $\ket{\psi_t}$, \hskip 1in 
{\bf Output}: An estimate of $\theta_t$. \\
Note: $\ket{r_1}$ ($\ket{r_2}$) is the state of the first register
     (second register, respectively). \\
(1) Let $\ket{r_1}\ket{r_2} = \ket{0}\ket{\psi_t}$. \\
(2) $\ket{r_1}\ket{r_2} = (H^{\otimes m}\otimes I) \ket{r_1}\ket{r_2}$. \\
(3) $\ket{r_1}\ket{r_2} = c_m(U) \ket{r_1}\ket{r_2}$. \\
(4) $\ket{r_1}\ket{r_2} = (QFT^{-1}\otimes I) \ket{r_1}\ket{r_2}$. \\
(5) $j$ = the result of measuring $\ket{r_1}$ \\
(6) Output $j/2^m$.
}

Step {\tt (1)} is an initialization of the registers into the state
$\ket{0}\ket{\psi_t}$ with input $\ket{\psi_t}$. Step {\tt (2)} applies the
Hadamard transformation to the first register, leaving the
registers in the state
\begin{equation}
   \ket{r_1}\ket{r_2}={\onesq{2^m}}\sum_{k=0}^{2^m-1}\ket{k}\ket{\psi_t}.
\end{equation}
As a result of applying $c_m(U)$ in step {\tt (3)}, the registers are in the
state
\begin{eqnarray}
   \ket{r_1}\ket{r_2} 
%   &=& c_m(U)({\onesq{2^m}}\sum_{k=0}^{2^m-1} \ket{k}\ket{\psi_t}) \nonumber \\
%   &=& {\onesq{2^m}}\sum_{k=0}^{2^m-1} \ket{k}U^k\ket{\psi_t} \nonumber \\
   &=& {\onesq{2^m}}\sum_{k=0}^{2^m-1} e^{2\pi \imath k \theta_t}
       \ket{k}\ket{\psi_t}. \label{eqn:ctrlstep}
\end{eqnarray}
To understand the workings of step {\tt (4)}, let us assume that $\theta_t =
j/2^m$, for some $j \in \setof{0,\ldots,2^m-1}$. Therefore, the equation
(\ref{eqn:ctrlstep}) can be rewritten as:
\begin{equation}
   \ket{r_1}\ket{r_2}={\onesq{2^m}}\sum_{k=0}^{2^m-1} e^{2\pi\imath k j/2^m}
                      \ket{k}\ket{\psi_t}.
\end{equation}
By applying the inverse quantum Fourier transform in step {\tt (4)}, the
registers are in the state
\begin{equation}
   \ket{r_1}\ket{r_2}= \ket{j}\ket{\psi_t}.
\end{equation}
By making a measurement on the first register in step {\tt (5)}, we obtain $j$,
where $\theta_t = j/2^m$.

\begin{figure}[h]
\begin{center}
   \includegraphics{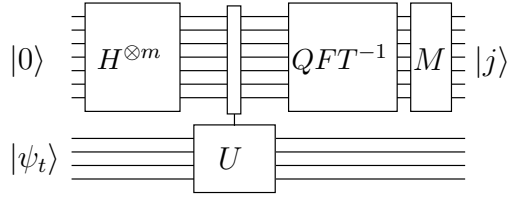}
   \caption{\label{fig:phaseest} 
      This figure shows the construction of a phase estimation circuit. The
      $m$-control $U$, $c_m(U)$, is not shown in detail.  However, if we have
      access to $U^{2^i}$, where $i \in\setof{0,1,\ldots}$, then $c_m(U)$ can
      be implemented using the method of repeated squaring. As a result, 
      $j/2^m$ is the best $m$-bit estimation of $\theta_t$.
   }
\end{center}
\end{figure}

In general, $\theta_t$ may not be of the form of $j/2^m$. However, the result
of applying the inverse QFT in step {\tt (4)} results in $j/2^m$ being the best
$m$-bit estimation of $\theta_t$ with a probability of at least $4/\pi^2$. For
more details, please consult Cleve et al~\cite{cleve:qal-revisited}. A quantum
circuit of the phase estimation algorithm is shown in figure
\ref{fig:phaseest}.

%Unfortunately, we do not know how to construct $c_m(U)$, in general, without
%any knowledge about $U$. If we have access to $U^{2^i}$, where $i \in
%\setof{0,1,2,\ldots}$, we can construct $c_m(U)$ by using the method of
%repeated squaring.

\subsection{Order Finding Algorithm}
%%%%%%%%%%%%%%%%%%%%%%%%%%%%%%%%%%%%%%%%%%%%%%%%%%%%%%%%%%%%%%%%%%%%%%%%%%%%%
{\bf The order finding problem} is defined as follows: {\em Given a positive
integer $N$ and an integer $a$ relatively prime to $N$, find the smallest
positive integer $r$ such that }
\begin{equation}
   a^r = 1\ (mod\ N). 
\end{equation}
First of all, we want a unitary transformation to use in the phase estimation
algorithm. We call that unitary transformation $M_a$, which is defined as
follows: 
\begin{equation}
    M_a:\ket{x} \rightarrow \ket{ax\ (mod\ N)},
\end{equation}
where $\ket{x}$ is an $n$-qubit register (the second register). Let
$\omega=e^{2\pi\imath \over r}$, and for each $k\in\setof{0,\ldots,2^n-1}$,
define 
\begin{equation} 
   \ket{\psi_t} = \onesq{r}\sum_{s=0}^{r-1} \omega^{-st}\ket{a^s}.
\end{equation} 
Then, for each $t$, \( M_a\ket{\psi_t} = \omega^t\ket{\psi_t}. \) 
In other words, $\omega^t$ is an eigenvalue of $M_a$ with respect to
eigenvector $\ket{\psi_t}$. Furthermore, $\theta_t \approx t/r$, for each $t$.
Therefore, if we have given an eigenvector $\ket{\psi_1}$, and we know how to
construct $c_m(M_a)$, then we can find $r$ (which is the period of $a$) by
using the phase estimation algorithm.

Unfortunately, it is not trivial to construct $\ket{\psi_t}$ for every $t$.
Instead of using $\ket{\psi_t}$, we use $\ket{1}$ which is effectively
equivalent to selecting $\ket{\psi_t}$, where $t$ is randomly selected from
$\setof{0, \ldots, r-1}$.  Then, we use the phase estimation algorithm to
compute the value of $j/2^m$ which is the best $m$-bit estimate value of $t/r$.
We extract the value of $t/r$ by using the continued fraction algorithm. If $t$
and $r$ are relatively prime, then we get $r$, which is the period of $a$. The
output $r$ of the phase estimation algorithm can be tested by checking that
$a^r = 1 (mod N)$. If $r$ is not the period of $a$, then we can re-execute this
algorithm until $t$ is coprime to $r$, which occurs with high probability in
$O(\log\log N)$ rounds~\cite{cleve:qal-revisited}.

In the next section, we describe an implementation of $c_m(U)$. This
calculation is equivalent to the calculation that Shor uses in his factoring
algorithm, known as ``modular exponentiation.'' Another detailed implementation
of the modular exponentiation can be found in Beckman et al
\cite{beckman:eff-factor}.

\subsection{An implementation of modular exponentiation}
\label{sec:modexp}
%%%%%%%%%%%%%%%%%%%%%%%%%%%%%%%%%%%%%%%%%%%%%%%%%%%%%%%%%%%%%%%%%%%%%%%%%%
To complete the implementation of the order finding algorithm, we need to
construct the unitary transformation $c_m(M_a)$. We accomplish this by using
the method of repeated squaring.

\noindent Let  $k_{m-1}k_{m-2}\ldots k_1k_0$, be the binary expansion of the
contents of the first register, $\ket{k}$. It now follows that
\begin{equation}
   M_a^k = \prod_{i=0}^{m-1} M_a^{k_i 2^i} = \prod_{i=0}^{m-1}
   (M_a^{2^i})^{k_i}.
\end{equation}
Then, for each $i$, we can implement the term $(M_a^{2^i})^{k_i}$ as a
controlled $M_a^{2^i}$, where the control qubit is $\ket{k_i}$.

\begin{figure}[h]
\begin{center}
   \includegraphics{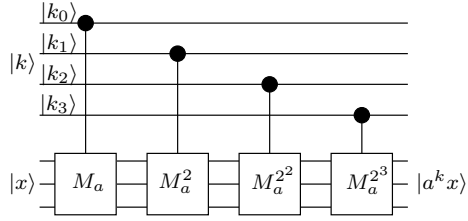}
   \caption{\label{fig:ctrl-mu} 
      Let $k_{m-1}\ldots k_1k_0$ be a binary representation of $k$. Then
      \(M_a^k = \prod_{i=0}^{m-1} M_a^{k_i 2^i}\). The term $M_a^{k_i 2^i}$
      is implemented as a control-$M_a^{2^i}$ circuit, where the control qubit
      is $\ket{k_i}$.
   }
\end{center}
\end{figure}

Please note that $a$ is a constant integer, and that $M_a^{2^i} = M_{a^{2^i}}$
for all $0 \le i < m$. Therefore, we can precompute the value of $a^{2^i}$ by
classical computers. Then we can apply the same technique used to implement
$M_a$, to implement $M_{a^{2^i}}$. Figure \ref{fig:ctrl-mu} shows an
implementation of $c_m(M_a)$.

\subsubsection{Reusing ancillary qubits}
\label{sec:revcomp}
%==============================================================================
For a given polynomial-time function $f$, we can construct a unitary
transformation $F$ which maps $\ket{x}\ket{0}$ to $\ket{x}\ket{f(x)}$. However,
the complete definition of $F$ also includes ancillary qubits which contain
information necessary for $F$ to be reversed. Let $g$ be a function that
computes the additional information, called ``garbage''. The complete definition
of $F$ is,
\begin{equation}
   F:\ket{x}\ket{0}\ket{0} \rightarrow \ket{x}\ket{f(x)}\ket{g(x)},
\end{equation}

The garbage needs to be reset, or erased, to state $\ket{0}$ before we make a
measurement.  Otherwise, the result of the measurement could be affected by the
garbage. To erase the garbage, Shor uses Bennett's technique which we review
in this section.

First we compute $F(x)$. Once we have the output $\ket{f(x)}$, we copy
$\ket{f(x)}$ into the extra register which has been preset to state $\ket{0}$.
Then we erase the output and the garbage of $F$ by reverse computing $F(x)$.
In particular, this procedure is described as follows:
\begin{eqnarray*}
   \ket{x}\ket{0}\ket{0}\ket{0} 
   & \stackrel{F}{\Longrightarrow} & \ket{x}\ket{f(x)}\ket{g(x)}\ket{0} \\ 
   & \stackrel{\COPY}{\Longrightarrow} & \ket{x}\ket{f(x)}\ket{g(x)}\ket{f(x)} \\
   & \stackrel{F^r}{\Longrightarrow} & \ket{x}\ket{0}\ket{0}\ket{f(x)},
\end{eqnarray*}
where $F^r$ is the reverse computation of $F$. We copy $\ket{f(x)}$ to the
extra register bit by bit by applying a $CNOT$ gate on each qubit. We define
$XF = F^r\cdot \COPY\cdot F$. 

If $f$ is a polynomial-time invertible function, we can create a unitary
transformation $OF$ which overwrites an input $\ket{x}$ with the output
$\ket{f(x)}$. We start from the construction of a unitary transformation $FI$
as follows: 
\begin{equation}
   FI:\ket{x}\ket{0} \rightarrow \ket{x}\ket{f^{-1}(x)},
\end{equation}
where $f^{-1}$ is a polynomial-time inverse function of $f$. The transformation
$FI$ may generate garbage, but it can be erased by using the technique
mentioned above. Finally, we implement $OF$ as follows:
\begin{eqnarray*}
   \ket{x}\ket{0}
   & \stackrel{F}{\Longrightarrow} & \ket{x}\ket{f(x)} \\ 
   & \stackrel{\SWAP}{\Longrightarrow} & \ket{f(x)}\ket{x} \\
   & \stackrel{FI^r}{\Longrightarrow} & \ket{f(x)}\ket{0},
\end{eqnarray*}
where $FI^r$ is the reverse computation of $FI$. The $\SWAP$ is a swap gate
that swaps the content of the input and the output registers. 
%For more information,
%please consult Shor~\cite{shor:factoring} or Beckman et
%al~\cite{beckman:eff-factor}.

\subsubsection{Binary adders}
\label{sec:binadder}
\begin{figure}[h]
\begin{center}
   \includegraphics{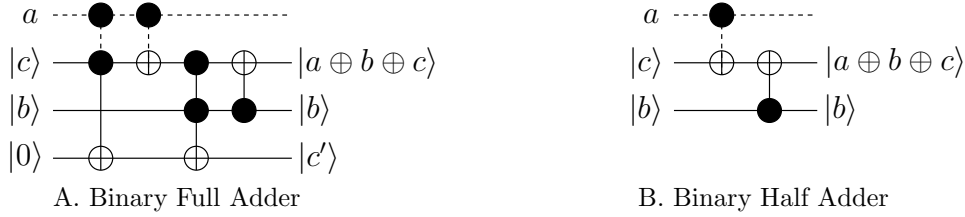}
   \caption{\label{fig:bfabha} 
      The dotted-line represents the classical bit $a$ which is used to
      control the quantum gates. The $\BFA_a$ computes an additional output
      carry qubit, while the $\BHA_a$ does not.
   }
\end{center}
\end{figure}

%==============================================================================
We continue our construction of $M_a$ by first implementing ``binary adders.''
There are two types of binary adders, ``binary full adder'' and ``binary half
adder,'' denoted by $\BFA_a$ and $\BHA_a$, respectively.  The $\BFA_a$ and
$\BHA_a$ are defined as follows: 
\begin{eqnarray}
   \BFA_a:\ket{c}\ket{b}\ket{0} \rightarrow \ket{a\oplus b\oplus c}
         \ket{b}\ket{c'} & &
   \BHA_a:\ket{c}\ket{b} \rightarrow \ket{a\oplus b\oplus c}\ket{b},
\end{eqnarray}
where $a$ is a classical bit, and $\ket{c}$ and $\ket{c'}$ are input and output
carries, respectively. The circuits for $\BFA_a$ and $\BHA_a$ are shown in
figure~\ref{fig:bfabha}.

The dotted-line represents a classical bit $a$ which is used to control the
quantum gates. If the classical bit is $0$, the quantum gate is not applied. If
the classical bit is $1$, then the quantum gate is applied.  The binary full
adder adds a classical bit $a$ to the carry $\ket{c}$ first, then adds a qubit
$\ket{b}$ to the sum. Because the carry is not computed by $\BHA_a$, we remove
two gates (the first gate, and the Toffoli gate) from $\BFA_a$ in order to
implement $\BHA_a$. 

\subsubsection{An $n$-qubit adder}
\label{sec:adder}
\begin{figure}[h]
\begin{center}
   \includegraphics{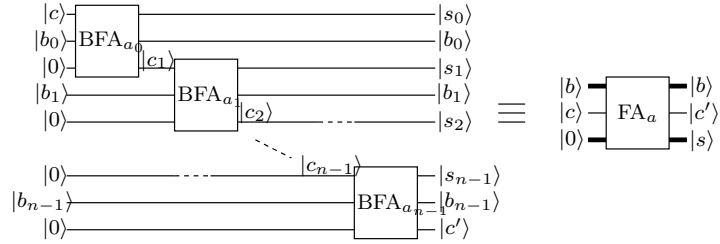}
   \caption{\label{fig:adders} 
      By applying the $\BFA$s bit by bit, as shown in the above circuit, we
      effectively add an $n$-bit number $a$ to the $n$-qubit number $\ket{b}$
      and the carry $\ket{c}$. The outputs are registers $\ket{b}$, $\ket{s}$,
      and a new carry $\ket{c'}$, where $s=a+b+c (mod\ 2^n)$. The thick lines
      represent an $n$-qubit register.
   }
\end{center}
\end{figure}

%==============================================================================
For each classical $n$-bit integer $a$, an $n$-qubit full adder $\FA_a$ is the
unitary transformation defined by 
\begin{equation}
   \FA_a:\ket{b}\ket{0}\ket{c} \rightarrow \ket{b}\ket{s}\ket{c'}
\end{equation}
where $\ket{s}$ is an $n$-qubit register, $s=a+b+c\ (mod\ 2^n)$, and $c$ and
$c'$ are an input and output carries, respectively. A quantum circuit for
$\FA_a$ is shown in figure~\ref{fig:adders}, where $a_{n-1}\cdots a_1a_0$,
$b_{n-1}\cdots b_1b_0$, and $c_{n-1}\cdots c_1c_0$ are $n$-bit binary
representations of $a, b$ and $c$, respectively. 

We replace the last $\BFA_{a_{n-1}}$ with a $\BHA_{a_{n-1}}$ to construct
$HA_a$.  As a result, we need only $n-1$ input ancillary qubits with initial
state $\ket{0}$ to implement $\HA_a$. By including an input carry qubit
$\ket{c}$, the $\HA_a$ is a $2n$-qubit unitary transformation.

\subsubsection*{An $n$-qubit adder modulo $N$}
\begin{figure}[h]
\begin{center}
   \includegraphics{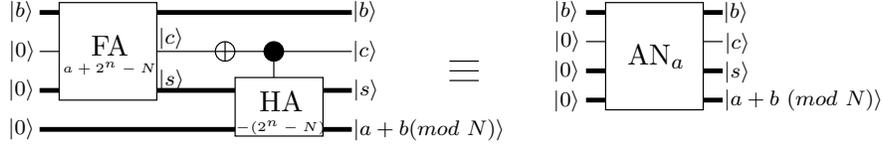}
   \caption{\label{fig:addmodn} 
      First, we add a number $a+2^n-N$ to ket $\ket{b}$. If the carry bit is
      not set, then we subtract $2^n-N$ from the sum. As a result, we compute
      $a+b\ (mod\ N)$.
   }
\end{center}
\end{figure}

%------------------------------------------------------------------------------
We use $\FA_a$ and $\HA_a$ to implement the $n$-qubit adder modulo $N$,
$(\AN_a)$.  We observe that if $a+b<N$, then $a+b\ (mod\ N)=a+b\ (mod\ 2^n)$;
otherwise $a+b\ (mod\ N) = a+b+2^n-N\ (mod\ 2^n)$.  We implement $\AN_a$ as
follows: First we compute the sum of $\ket{b}$ with a classical number
$a+2^n-N$ in modulo $2^n$. If the carry is not set, then we subtract $2^n-N$
from the sum. Hence, we have a transformation $\AN_a$, given by
\begin{equation}
   \AN_a:\ket{b}\ket{0}\ket{0}\ket{0} \rightarrow \ket{b}\ket{s}\ket{c}\ket{a+b\
   (mod\ N)}.  
\end{equation}
where $s=b + a + 2^n - N \ (mod\ 2^n)$, and $c$ is the carry.  The circuit that
implements $\AN_a$ is shown in figure~\ref{fig:addmodn}.

\begin{figure}[h]
\begin{center}
   \includegraphics{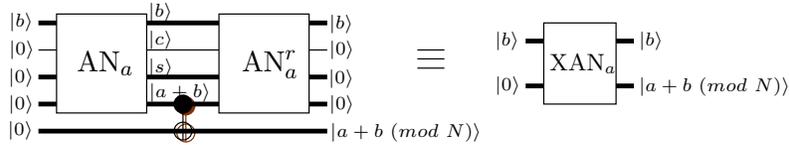}
   \caption{\label{fig:xaddmodn} 
      Using a $CNOT$ gate to copy bit by bit from the output register of $\AN_a$
      to an $n$-qubit ancillary register, we can apply $\AN_a^r$ so that
      $\ket{s}$ and $\ket{c}$ are set to state $\ket{0}$.  
   }
\end{center}
\end{figure}

We use the technique described in section $\ref{sec:revcomp}$ to reset
$\ket{s}$ and $\ket{c}$ back to state $\ket{0}$. As a result, we obtain a
transformation $\XAN_a = \AN_a^r \cdot \COPY \cdot \AN_a$ which acts as
follows:
\begin{equation}
   \XAN_a:\ket{b}\ket{0}\ket{0}\ket{0}\ket{0} \rightarrow
   \ket{b}\ket{0}\ket{0}\ket{0}\ket{a+b\ (mod\ N)}.  
\end{equation}
In other words, $\XAN_a$ is a $2n$-qubit transformation (with $2n+1$ ancillary
qubits) which sends $\ket{b}\ket{0}$ to $\ket{b}\ket{a+b\ (mod\ N)}$. The wiring
diagram for $\XAN_a$ is shown in figure \ref{fig:xaddmodn}.

Because the inverse transformation of $\XAN_a$ is $\XAN_{-a}$, the input of
$\XAN_a$ can be overwritten by the output of $\XAN_a$ by using the technique
described in section $\ref{sec:revcomp}$.  We now define the adder $A_a$ as
follows:
\begin{equation}
   A_a = \XAN_{-a}^r \cdot \SWAP \cdot \XAN_{a}. 
\end{equation} 
As a result, the transformation $A_a$ is an $n$-qubit transformation (with
$3n+1$ ancillary qubits) which maps $\ket{b}$ to $\ket{a+b\ (mod\ N)}$.

\subsubsection{An $n$-qubit multiplier}
\label{sec:multi}
%==============================================================================
Now we are ready to describe the construction of $M_a$, which maps $\ket{x}$ to
$\ket{ax}$, where $x \in \intset_N$. We define an $2n$-qubit unitary
transformation $MF_a$ as follows: 
\begin{equation}
   MF_a:\ket{x}\ket{0}\rightarrow \ket{x}\ket{ax (mod\ N)}.
\end{equation}
Assuming $x_{n-1}\ldots x_1x_0$ is the binary representation of $x$, we have
\begin{equation}
   ax = \sum_{i=0}^{n-1} a x_i2^i.
\end{equation}
For each $0 \le i < n$, the term $a x_i
2^i$ can be implemented by the control-$A_{a2^i}$, where $\ket{x_i}$ is a
control qubit, and 
\begin{equation}
   A_{a2^i}:\ket{b} \rightarrow \ket{b+a2^i\ (mod\ N)}. 
\end{equation}
Since for each $i$, $a2^i$ is constant, we can compute each $a2^i$ by using a
classical computer. Then we use the result and the same technique for
implementing $A_a$, as described in section \ref{sec:adder}, to construct
$A_{a2^i}$.  Therefore, the transformation $MF_a$ can be implemented using the
method of repeated squaring with a circuit similar to the circuit shown in
figure~\ref{fig:ctrl-mu}.  Hence, $MF_a$ is a $2n$-qubit transformation sending
$\ket{x}\ket{0}$ to $\ket{x}\ket{ax}$, using of $3n+1$ ancillary qubits.

Finally, with the overwriting output technique described in section
\ref{sec:revcomp}, the transformation $M_a$ can be implemented as $M_a =
MF_{a^{-1}}\cdot \SWAP\cdot MF_a$. (Note that, because $a$ and $N$ are
relatively prime, $a^{-1}$ always exists in $\intset_N$.) In other words, $M_a$
is an $n$-qubit transformation with $4n+1$ ancillary qubits. Thus, the so
constructed $M_a$ can be plugged into the transformation $c_m(M_a)$, as
described earlier in section \ref{sec:modexp}.

\subsection{Complexity analysis}
%%%%%%%%%%%%%%%%%%%%%%%%%%%%%%%%%%%%%%%%%%%%%%%%%%%%%%%%%%%%%%%%%%%%%%%%%%
We analyze the complexity of our implementation of Shor's algorithm for two
parameters,i.e., the number of gates and the number of qubits.

\subsubsection*{Gate complexity}
%=========================================================================
To count the number of gates, we define a function $G(F)$ to be the number of
gates used to implement the transformation $F$. We recursively compute the
number of gates as follows:
\begin{eqnarray*}
   G(SHOR) & = & G(H^{\otimes m} + G(c_m(M_a)) + G(QFT^{-1}) \\
   G(c_m(M_a)) & = & m\cdot G(M_a) \\
   G(M_a) & = & n\cdot G(A_a) \\
   G(A_a) & = & 2\cdot G(\XAN_a) + G(\SWAP) \\
   G(\XAN_a) & = & 2\cdot G(\AN_a) + G(\COPY) \\
   G(\AN_a) & = & G(\FA_a) + G(\HA_a) + 1 \\
   G(\FA_a) & = & n\cdot G(\BFA) = 4n \\
   G(\HA_a) & = & (n-1)\cdot G(\BFA) + G(\BHA) = 4n-2
\end{eqnarray*}
Since $G(H^{\otimes m})=m$, $G(QFT^{-1})=m(m+1)/2=O(m^2)$, and
$G(c_m(M_a))=70mn^2-6mn = O(mn^2)$, it follows that the gate complexity of this
implementation is $O(mn^2)$. In general, $m=2n$. Therefore, the complexity is
$O(n^3)$.

However, we count a control-gate with multiple control-qubits as one gate.  In
fact, a control gate with multiple control qubit can be broken down into a
sequence of Toffoli gates using the techniques described by Beranco et
al~\cite{barenco:elemgate}.  Moreover, the number of needed Toffoli gates grows
exponentially with respect to the number of control qubits in the control-gate.
Fortunately, the number of control qubits in the Shor's algorithm is at most
$5$: One control qubit for $c_m(M_a)$, one control qubit for $M_a$, one control
qubit for control-$\FA_a$ in the implementation of $\AN_a$, and two control
qubits in the implementation of $\FA_a$. Moreover, the number of control qubits
does not depend on the input number $N$.  Therefore, there is constant
overhead from breaking down a control gate with multiple control qubits into a
sequence of Toffoli gates. This overhead does not have affect the gate
complexity.

\subsubsection*{Space complexity}
First of all, $\XAN_a$ is a $2n$-qubit transformation with $2n+1$ ancillary
qubits.  So, we need $n$ qubits to control the transformation $A_a$ in the
implementation of $M_a$, and $m$ more qubits to control the transformation
$M_a$ in the implementation of $c_m(M_a)$.  Therefore, the number of qubits
needed in this implementation is $5n+m+1$.

\section{Distributed quantum factoring algorithm}
\label{sec:distshor}
%%%%%%%%%%%%%%%%%%%%%%%%%%%%%%%%%%%%%%%%%%%%%%%%%%%%%%%%%%%%%%%%%%%%%%%%%%%%%%
We implement a distributed quantum factoring algorithm as briefly described as
follows: First, we partition $5n+m+1$ qubits into groups in such a way that
each group fits on one of the quantum computers making up a network. Then, we
implement a distributed quantum factoring algorithm on this quantum network by
replacing a control gate with a non-local control gate, whenever necessary.  

In this paper, we will describe a distributed quantum factoring algorithm to
factor a number $N$ within specific parameters. We assume that we have a
network of $(n+c)$-qubit quantum computers, where $n=\log N$. The $c$ extra
qubits for each computer can be used as either channel qubits or ancillary
qubits. We will show that $c$ is a constant which does not depend on the input
number $N$.  To be more specific, we choose $m=2n$.  Therefore, the number of
qubits needed in this implementation is $7n+1$ qubits. Although, this
particular implementation is specific to certain parameters, its implementation
can easily be generalized. 

First we divide the control register of $c_m(M_a)$, $\ket{k}$, into two
$n$-qubits groups.  Then we place these two groups on two different computers.
Each qubit $\ket{k_i}$ of these two groups remotely controls the transformation
$M_a^{2^i}$.

Another computer is assigned to hold the control register of $MF_a$, i.e.,
$\ket{x}$.  Each qubit $\ket{x_j}$ remotely controls the transformation $A_a$.

\begin{figure}[h]
\begin{center}
   \includegraphics{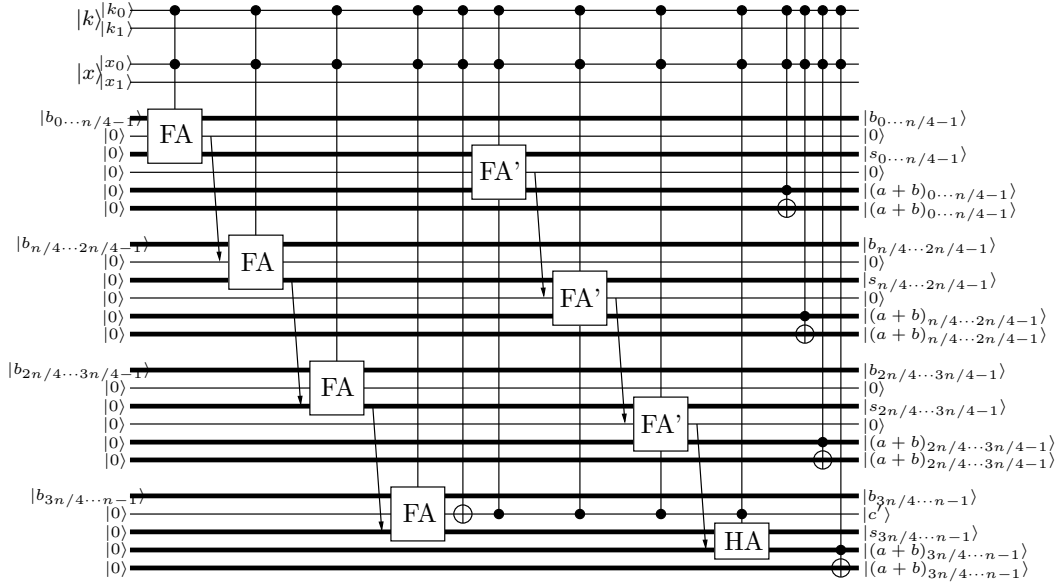}
   \caption{\label{fig:distadder} 
      This figure shows how to compute $\AN_a$ followed by $\COPY$
      transformations. Each computer holds $n/4$ qubit from each register.
      Each computer also has two carry qubits which have been set to $\ket{0}$.
      The arrow line represents teleportation of the output carry qubit to the
      next computer. Each transformation is remotely controlled by two qubits,
      one from register $\ket{k}$, and the other from the register $\ket{x}$.  
   }
\end{center}
\end{figure}

Next, we implement the transformation $\XAN_a$, which is a component of
$\AN_a$.  The transformation $\XAN_a$ has two registers, one $n$-qubit input
register $\ket{b}$, and one $n$-qubit output register $\ket{a+b}$. However,
$\XAN_a$ also requires $2n+1$ ancillary qubits, i.e., one carry bit, $n$ qubits
for the intermediate sum $\ket{s}$, and $n$ qubits for the intermediate output
register $\ket{a+b}$. Therefore, it takes four computers to compute $\XAN_a$.
Each computer computes $1/4$ of each register, as shown in figure
\ref{fig:distadder}. Each computer holds $n/4$ qubits from the input registers
$\ket{b}$, $n/4$ qubits from the intermediate sum register $\ket{s}$, $n/4$
qubits from the intermediate output register, and $n/4$ qubits from the output
register $\ket{a+b}$ (represented by thick lines). Each computer also has two
extra carries qubits, which are used in computing of $\FA$ and $\FA'$.

\begin{figure}[h]
\begin{center}
   \includegraphics{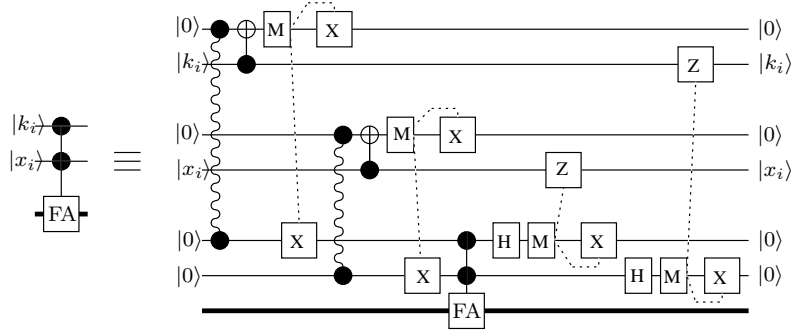}
   \caption{\label{fig:mctrl-fa} 
      A distributed multi-control gate can be implemented by distributing all
      control qubits to the computer that hold the target qubits, then
      implementing the multi-control gate locally. This figure shows how to
      implement an distributed control-control-FA gate.
   }
\end{center}
\end{figure}

The first four $\FA$ transformations compute $\FA_{a+2^n-N}$ with the input
carry $\ket{c} = \ket{0}$.  These $\FA$s are remotely controlled by two control
qubits, one from the register $\ket{k}$, and the other from the register
$\ket{x}$.  A distributed control $\FA$ with two control qubits is implemented
by distributing two control qubits onto the computer that holds the target
qubits, and then implementing the double control locally, as shown in
figure~\ref{fig:mctrl-fa}.  After completing each $\FA$ computation, the output
carry bit is teleported to the next $\FA$ on another computer. The
teleportations are represented by arrow lines. 

The transformation $\HA_{-(2^n-N)}$ is computed by the next three full adders
$\FA'$, and a half adder $\HA$. The integer $-(2^n-N)$ is precomputed by a
classical computer, and then used to implement $\FA'$ and $\HA$.  Similarly,
the carry qubit is teleported from one computer to another. The last carry
qubit is teleported into the first qubit of the intermediate output register
$\ket{(a+b)_{3n/4\cdots n-1}}$.  

Each $\FA'$ and the single $\HA$ are each controlled by three qubits: One from
the first register $\ket{k}$, one from the register $\ket{x}$, and the last
from the output carry bit of $\FA_{a+a^n-N}$. A non-local three-control gate
can implemented by distributing all three-control qubits onto the target
computer, and locally implementing the control gate with three control qubits.

The $\COPY$ transformation is a bitwise copy implemented in terms of CNOT
gates. Because each computer possesses $n/4$ of intermediate output register
and the final output register itself, the distributed $\COPY$ can be easily
implement by locally applying CNOT gates, as shown in figure
\ref{fig:distadder}.  However, $\COPY$ still needs to be remotely controlled by
two qubits from register $\ket{k}$ and register $\ket{x}$.  

Similarly, each machine possesses $n/4$ qubits of both input register and
output register. The distributed $\SWAP$ can be locally implement on each
machine, remotely controlled by two qubits from register $\ket{k}$ and
register $\ket{x}$.  

\subsubsection*{The number of extra qubits}
%-------------------------------------------------------------------------
The number of extra qubits $c$ depends on two factors: The number of channel
qubits, and the number of extra carry qubits needed in the implementation. The
number of channel qubits depends on how many non-local control qubits are
needed. In this implementation, at most three non-local control qubits are
implemented. Therefore, at most $3$ channel qubits are required at one time.
Furthermore, there are only two extra carry qubits (one carry qubit for
transformation $\FA$ and another carry qubit for transformation $\FA'$) needed
in this implementation.  Therefore, $c = 5$, and does not depend on the input
$N$.

\subsection{Communication complexity}
%=========================================================================
By communication complexity, we means the number of entangled pairs needed to
be established, and the number of classical bits needed to be transmitted in
each direction. The optimum cost of implementing a non-local operation is one
EPR pair and two classical bits (one in each direction). Therefore, if we can
count the number of non-local control gates and teleportation circuits, we can
estimate the communication overhead.  The communication overhead of a control
gate with multiple control qubits (such as control-FA with two control qubits)
is equal to the overhead for a single non-local CNOT gate multiplied by the
number of control qubits. Fortunately, the maximum number of non-local control
qubits is at most $3$. Therefore, we can count every gate as one control gate.

If we simply count every gate as a non-local gate, the communication overhead
is $O(mn^2)$. This number is an over estimation because the cost of each
non-local control-$U$ gate, where $U$ can be decomposed into a number of
elementary gates, can be shared among these elementary gates.

To be more precise, we define a function $NL(F)$ to be the number of non-local
control gates implemented in the distributed implementation of circuit $F$. We
compute $NL(SHOR)$ as follows: 
\begin{eqnarray*}
   NL(SHOR) & = & NL(H^{\otimes m} + NL(c_m(M_a)) + NL(QFT^{-1}) \\
   NL(c_m(M_a)) & = & m\cdot NL(M_a) \\
   NL(M_a) & = & n\cdot NL(A_a) \\
   NL(A_a) & = & 2\cdot NL(\XAN_a) + NL(\SWAP) \\
   NL(\XAN_a) & = & 2\cdot NL(\AN_a) + NL(\COPY).
\end{eqnarray*}

As shown in figure~\ref{fig:distadder}, there are $8$ non-local control gates
per $\AN_a$, i.e., $NL(\AN_a) = 8$. (The non-local control NOT gate in the
middle can be included in the implementation of the last non-local control
$\FA$.) Four non-local control circuits are sufficient to implement $\COPY$.
Similarly, another four non-local control circuits are sufficient to implement
$\SWAP$.  Therefore, $NL(c_m(M_a)) = 44mn = O(mn)$.  Since $NL(QFT^{-1}) =
O(m^2)$, then $NL(SHOR) = O(mn + m^2)$. 

Similarly, we define a function $T(F)$ to be the number of teleportation
circuits implemented in the distributed implementation of circuit $F$. Then, six
teleportation circuits are sufficient to implement $\AN_a$. There is no need
for a teleportation circuit in $\COPY$, $\SWAP$, and $QFT^{-1}$. Therefore,
$T(SHOR) = 12mn = O(mn)$.  

As a result, the communication complexity of Shor is $NL(SHOR)+T(SHOR) = O(mn +
m^2)$. In this particular implementation, $m=2n$. Hence, the communication over
is $O(n^2)$.

\section{Acknowledgments}
%%%%%%%%%%%%%%%%%%%%%%%%%%%%%%%%%%%%%%%%%%%%%%%%%%%%%%%%%%%%%%%%%%%%%%%%%%%%%%
This effort is partially supported by the Defense Advanced Research Projects
Agency (DARPA) and Air Force Research Laboratory, Air Force Materiel Command,
USAF, under agreement number F30602-01-2-0522, the National Institute for
Standards and Technology (NIST). The U.S. Government is authorized to reproduce
and distribute reprints for Government purposes notwithstanding any copyright
annotations thereon. The views and conclusions contained herein are those of
the authors and should not be interpreted as necessarily representing the
official policies or endorsements, either expressed or implied, of the Defense
Advanced Research Projects Agency, the Air Force Research Laboratory, or the
U.S. Government.

\end{document}